\documentclass{article}
\usepackage{amsmath, amssymb, amsfonts}
\title{Semiclassical corrections to a regularized Schwarzschild metric} 
\author{Hristu Culetu, \\Ovidius University, Dept.of Physics and Electronics, \\ Mamaia Avenue 124, 900527 Constanta, Romania, \\e-mail : hculetu@yahoo.com}

\begin{document}
\numberwithin{equation}{section}
\pagenumbering{arabic}
\maketitle
\newcommand{\fv}{\boldsymbol{f}}
\newcommand{\tv}{\boldsymbol{t}}
\newcommand{\gv}{\boldsymbol{g}}
\newcommand{\OV}{\boldsymbol{O}}
\newcommand{\wv}{\boldsymbol{w}}
\newcommand{\WV}{\boldsymbol{W}}
\newcommand{\NV}{\boldsymbol{N}}
\newcommand{\hv}{\boldsymbol{h}}
\newcommand{\yv}{\boldsymbol{y}}
\newcommand{\RE}{\textrm{Re}}
\newcommand{\IM}{\textrm{Im}}
\newcommand{\rot}{\textrm{rot}}
\newcommand{\dv}{\boldsymbol{d}}
\newcommand{\grad}{\textrm{grad}}
\newcommand{\Tr}{\textrm{Tr}}
\newcommand{\ua}{\uparrow}
\newcommand{\da}{\downarrow}
\newcommand{\ct}{\textrm{const}}
\newcommand{\xv}{\boldsymbol{x}}
\newcommand{\mv}{\boldsymbol{m}}
\newcommand{\rv}{\boldsymbol{r}}
\newcommand{\kv}{\boldsymbol{k}}
\newcommand{\VE}{\boldsymbol{V}}
\newcommand{\sv}{\boldsymbol{s}}
\newcommand{\RV}{\boldsymbol{R}}
\newcommand{\pv}{\boldsymbol{p}}
\newcommand{\PV}{\boldsymbol{P}}
\newcommand{\EV}{\boldsymbol{E}}
\newcommand{\DV}{\boldsymbol{D}}
\newcommand{\BV}{\boldsymbol{B}}
\newcommand{\HV}{\boldsymbol{H}}
\newcommand{\MV}{\boldsymbol{M}}
\newcommand{\be}{\begin{equation}}
\newcommand{\ee}{\end{equation}}
\newcommand{\ba}{\begin{eqnarray}}
\newcommand{\ea}{\end{eqnarray}}
\newcommand{\bq}{\begin{eqnarray*}}
\newcommand{\eq}{\end{eqnarray*}}
\newcommand{\pa}{\partial}
\newcommand{\f}{\frac}
\newcommand{\FV}{\boldsymbol{F}}
\newcommand{\ve}{\boldsymbol{v}}
\newcommand{\AV}{\boldsymbol{A}}
\newcommand{\jv}{\boldsymbol{j}}
\newcommand{\LV}{\boldsymbol{L}}
\newcommand{\SV}{\boldsymbol{S}}
\newcommand{\av}{\boldsymbol{a}}
\newcommand{\qv}{\boldsymbol{q}}
\newcommand{\QV}{\boldsymbol{Q}}
\newcommand{\ev}{\boldsymbol{e}}
\newcommand{\uv}{\boldsymbol{u}}
\newcommand{\KV}{\boldsymbol{K}}
\newcommand{\ro}{\boldsymbol{\rho}}
\newcommand{\si}{\boldsymbol{\sigma}}
\newcommand{\thv}{\boldsymbol{\theta}}
\newcommand{\bv}{\boldsymbol{b}}
\newcommand{\JV}{\boldsymbol{J}}
\newcommand{\nv}{\boldsymbol{n}}
\newcommand{\lv}{\boldsymbol{l}}
\newcommand{\om}{\boldsymbol{\omega}}
\newcommand{\Om}{\boldsymbol{\Omega}}
\newcommand{\Piv}{\boldsymbol{\Pi}}
\newcommand{\UV}{\boldsymbol{U}}
\newcommand{\iv}{\boldsymbol{i}}
\newcommand{\nuv}{\boldsymbol{\nu}}
\newcommand{\muv}{\boldsymbol{\mu}}
\newcommand{\lm}{\boldsymbol{\lambda}}
\newcommand{\Lm}{\boldsymbol{\Lambda}}
\newcommand{\opsi}{\overline{\psi}}
\renewcommand{\tan}{\textrm{tg}}
\renewcommand{\cot}{\textrm{ctg}}
\renewcommand{\sinh}{\textrm{sh}}
\renewcommand{\cosh}{\textrm{ch}}
\renewcommand{\tanh}{\textrm{th}}
\renewcommand{\coth}{\textrm{cth}}

\begin{abstract}
A version of the Schwarzschild metric to be valid in microphysics is proposed. The source fluid is anisotropic with $p_{r} = -\rho$ and fluctuating tangential pressures. At large distances with respect to the Compton wavelength associated to the source particle, they do not depend on the mass $m$ of the source and everywhere depend on $\hbar$ and the velocity of light $c$ but not on the Newton constant $G$. The particle may be a black hole for $m \geq m_{P}$ only and when $m = m_{P}$ becomes an extremal black hole. The Komar energy $W$ of the gravitational fluid is $mc^{2}$ for $\hbar = 0$ and at large distances and vanishes at $r_{0} = 2\hbar/emc$. The WEC is violated when $r < r_{0}/2$ due to the negative tangential pressures. The horizon entropy for the extremal black hole is finite though $W$ and the temperature $T$ are vanishing there.
 \end{abstract}
 
 \section{Introduction}
 The singularities which were predicted to form inside black holes (BHs) are generally regarded as indicating the breakdown of the General Relativity, requiring modifications of the theory, including Quantum Mechanics and possibly Quantum Gravity (QG). Several models for non-singular, static, spherically-symmetric BHs have been considered so far \cite{JB, ID, BR, SH, PN, HC1, LPRS, RV, BRF, JN}. Bonanno and Reuter \cite{BR} altered the Schwarzschild (KS) metric using a running Newton constant and studied the quantum gravitational effects on the dynamics of geometry. In their view, the black hole evaporation stops when its mass approaches some extremal (critical) value $M_{cr} \approx m_{P}$, where $m_{P}$ is the Planck mass. Consequently, a ''cold'' soliton-like remnant is formed so that the classical singularity at $r = 0$ is removed. According to the authors of \cite{BR}, that quantum BH with $M = M_{cr}$ (corresponding to the extremal charged Reissner - Nordstrom BH) may be considered the final state of the KS black hole evaporation process. 
	
	Hayward \cite{SH} introduced the so-called ''regular'' black holes (RBHs) which avoid the curvature singularity beyond the event horizon. The singularity is replaced by a de Sitter spacetime. Nicolini \cite{PN} studied the relation between quantum BHs and Noncommutative Geometry to cure the singular behavior of gravity at the BH center, replacing it with a stable remnant. Rovelli and Vidotto \cite{RV} suggested that QG becomes relevant when the energy density of matter reaches the Planck value, which may happen at length scales much larger than planckian. A star that collapses may reach a new stage where quantum gravitational pressure counteracts the matter's weight. The collapse ends when matter bounces out and the star can avoid sinking into $r = 0$ singularity. That process satisfies the Einstein equations everywhere except for a small region where quantum effects dominate. 
	
		We propose in this work a simple modification of the Schwarzschild metric which renders it regular everywhere, in the spirit of \cite{HC1}. Nevertheless, we are now interested to apply the model in microphysics, so that the appearance of the Planck constant is unavoidable. Therefore, small or large distances are compared not with respect to the horizon radius but with respect to the Compton wavelength $\lambda_{c}$ of the source particle. The field becomes repulsive for $r < \lambda_{c}$ as it was previously suggested by De Lorenzo et al. \cite{LPRS}, Hayward \cite{SH} and Barrau and Rovelli \cite{BRF}. The red-shift factor $f = -g_{tt}$ has a similar behaviour with $F(r)$ of Hayward and De Lorenzo et al.. We are specially interested of the extremal case $m = m_{crit}$ when there is one horizon only. Due to the exponential factor in the metric, all quantities are finite at $r = 0$ and at infinity. Because of the very large tangential pressures the WEC is not obeyed, as De Lorenzo et al. already noticed in \cite{LPRS}. 
	
	Throughout the paper geometrical units $G = c = \hbar = k_{B} = 1$ are used, unless otherwise specified.
	
	\section{Regularized Schwarzschild metric}
	 Our purpose in this paper is to look for a non-singular KS-type spacetime. We firstly consider the Xiang et al. \cite{XLS} modified form of the KS line element, with $-g_{tt} = 1 - (2m/r)exp(-\alpha/r^{2})$, $\alpha$ being a length squared. Their choice removes the standard KS singularity at $r = 0$. However, at large distances the RN metric or the extremal BH metric are not retrieved when a second order radial power expansion of $-g_{tt}$ is performed
	     \begin{equation}
   -g_{tt} = 1 - \frac{2m}{r} + \frac{2\alpha m}{r^{3}} - ... ,
 \label{2.1}
 \end{equation} 
 when $\alpha$ is related to the BH charge.
	 
 To get rid of these inadequacies, we choose the red-shift factor as \cite{HC4} \footnote{This red-shift factor, with $k = q^{2}/2$, has recently been used by Rodrigues et al. \cite{RJMZ} as a ''new'' regular solution.}
     \begin{equation}
   f(r) = 1 - \frac{2m}{r} e^{-\frac{k}{mr}}   
 \label{2.2}
 \end{equation} 
 (see also \cite{HC1, BV}), where $m$ is the BH mass and $k$ is a positive dimensionless constant. In other words, $m$ has units $1/length$, i.e. the Planck constant will arise through the reduced Compton wavelength. That choice is suitable in microphysics where we are going to apply our model. The modified KS metric reads
	     \begin{equation}
   ds^{2} = -\left(1 - \frac{2m}{r} e^{-\frac{k}{mr}}\right) dt^{2} + \frac{1}{1 - \frac{2m}{r} e^{-\frac{k}{mr}}} dr^{2} + r^{2} d \Omega^{2},     
 \label{2.3}
 \end{equation}  
where $d \Omega^{2}$ stands for the metric on the unit 2-sphere. The first derivative of $f(r)$ appears as
 \begin{equation}
f'(r) = \frac{2m}{r^{2}} (1 - \frac{k}{mr}) e^{-\frac{k}{mr}}.
 \label{2.4}
 \end{equation}  
We see that $f'(r)$ is negative for $r < k/m$, positive for $r > k/m$ and zero for $r = k/m$. The metric function becomes minimal at $r = k/m$ where it takes the value 
 \begin{equation}
f_{min}(r) = 1 - \frac{2m^{2}}{ekm_{P}^{2}},
 \label{2.5}
 \end{equation}  
with $m_{P}$ the Planck mass and $lne = 1$. For simplicity and to avoid unimportant numerical factors we choose $k = 2/e$ and so the sign of $f_{min}$ will depend on the value of $m$ w.r.t. $m_{P}$. We distinguish three situations of interest here:\\
(1) $m < m_{P}$, when $f_{min} > 0$ and $f(r) = 0$ has no roots. The red-shift factor is positive for any $r$ and there is no any horizon.\\ 
(2) $m = m_{P}$, which gives $f_{min} = 0$. We have now a double root at $r_{H} = 2/me$ which represents the event horizon. Hence $f(r_{H}) = 0$ and $f'(r_{H}) = 0$ are simultaneously satisfied. The BH becomes extremal \cite{BR, HC3}, with a degenerate horizon. \\
(3) $m > m_{P}$, when $f_{min} < 0$. Equation $f(r) = 0$ has two roots: $r_{-} < r_{0} \equiv 2/me$ (the Cauchy horizon) and $r_{+} > r_{0}$ (the event horizon). However, their location cannot be determined analitically because of the transcendental nature of the equation. 

If the above model is realistic we may conclude that a particle with a mass less than the Planck mass cannot become a BH. That fits with the assumption that the Planck length $l_{P} = 10^{-33}$ cm is a minimal length. Indeed, a microparticle with $m < m_{P}$ has an unrealistic gravitational radius, eventually much less than $l_{P}$. With our previous value of $k$ the metric function $f(r)$ acquires the form
     \begin{equation}
   f(r) = 1 - \frac{2m}{r} e^{-\frac{2}{emr}}.   
 \label{2.6}
 \end{equation}  
A plot of $f(r)$ versus $r$ for a given mass looks similar with those of Hayward \cite{SH} and De Lorenzo et al. \cite{LPRS} (see also \cite{HC1}), even the discussion concerning the relation between $m$ and their $m^{*}$, which has been taken to be of the order of $m_{P}$. If we expand $f(r)$ for $r >> r_{0} = 2/em$, up to $r^{-2}$, one obtains
 \begin{equation}
   f(r) \approx 1 - \frac{2m}{r} + \frac{4l_{P}^{2}}{er^{2}}
 \label{2.7}
 \end{equation}  
and we see that the 3rd term on the r.h.s. of (2.7) no longer depends on the mass $m$. When we put above $\hbar = 0$, the standard KS metric is retrieved. Therefore, (2.3) seems to be a semiclassical expression of the KS spacetime. 

Let us now consider a static observer characterized by the velocity vector field $u^{b} = (1/\sqrt{f}, 0, 0, 0)$. The corresponding acceleration 4-vector is given by
  \begin{equation}
  a^{b} = \left(0, \frac{m(1 - \frac{r_{0}}{r})} {r^{2}} e^{-\frac{r_{0}}{r}}, 0, 0 \right).    
 \label{2.8}
 \end{equation} 
It is worth noting that the gravitational field becomes repulsive at $r < r_{0}$ (when $a^{r} < 0$) and vanishes at $r = r_{0}$. For $r >> r_{0}$, $a^{r}$ will no longer depend on the Planck constant and it acquires the Newtonian expression $m/r^{2}$. When $m = m_{P}$, $r_{0}$ represents the event horizon radius $r_{H}$ and the particle becomes an extremal BH. Therefore, its surface gravity will vanish (no Hawking radiation, as one should be for a degenerate horizon). We find that $r_{1} = r_{0} (2 - \sqrt{2})/2$ and $r_{2} = r_{0} (2 + \sqrt{2})/2$ are the locations of the minimum and, respectively maximum values of $a^{r}$. 

Let us take a numerical example and calculate the radial acceleration for an observer sitting in the gravitational field of, say, a neutron, with $m_{n} \approx 1.6 \cdot 10^{-27} Kg$ at a distance $r = r_{1}$ from the particle. One finds $a^{r} \approx -10^{-6} m/s^{2}$, where $e^{-\sqrt{2}} \approx 0.24$ has been used. $a^{r}$ is, of course, negative because $r = r_{1}$ is located in the repulsive core $r < r_{0}$. In other words, the geometry (2.6) leads to a repulsive gravity at very short distances, even though the accelerations are very tiny. If we consider a macroscopic value for $m$ (for instance, a solar mass star), the ratio $r_{0}/r$ is completely negligible even when the star radius approaches its gravitational radius. In this case we apply the standard KS geometry or eventually its modified forms from the cited papers.

\section{Source stress tensor}
 Let us find now what are the sources of the spacetime (2.3), with $k = 2/e$,  namely what energy-momentum tensor do we need on the r.h.s. of Einstein's equations $G_{ab} = 8\pi T_{ab}$ in order that (2.3) to be an exact solution. One finds that
   \begin{equation}
   \begin{split}
  T^{t}_{~t} = -\rho = - \frac{1}{2\pi er^{4}} e^{-\frac{2}{emr}},~~~ T^{r}_{~r} = p_{r} = - \rho,\\ T^{\theta}_{~\theta} = T^{\phi}_{~\phi} = p_{\theta} = p_{\phi} =  \frac{1}{2 \pi e r^{4}} \left(1 - \frac{1}{emr}\right) e^{-\frac{2}{emr}}.
  \end{split}
\label{3.1}
\end{equation}
We notice firstly that $\rho > p_{\theta}$ always and $p_{r} = - \rho$. Nevertheless, the fluid is anisotropic since $p_{r} \neq p_{\theta} = p_{\phi}$. The energy density and all pressures are non-singular at $r = 0$ and when $r \rightarrow \infty$ (where, actually, they vanish). Moreover, $\rho$ is positive for any $r$. The strong energy condition is not satisfied for $r < r_{0}/2$, where $\rho + \Sigma p_{i} = 2p_{\theta} < 0 ~(i = 1,2,3)$. The maximum value $\rho_{max} = 8m^{4}/\pi e$ is reached in the repulsive core, more precisely at $r = r_{0}/4$. For $r >> r_{0}$, $\rho(r)$ and all the pressures tends to zero. In addition, $\rho$ increases with $m$ at constant $r$ but it is independent on $m$ at large $r$ when the exponential factor might be neglected and $\rho \approx \hbar c/er^{4}$, a Casimir-type expression. Anyway, the energy density and pressures are all independent on the Newton constant $G$ but depend only on $c$ and $\hbar$. They turn out to have more a quantum than a gravitational origin in spite of the fact that we started with Einstein's equations. However, if we write $\rho$ for large $r$ as
   \begin{equation}
  \rho = G \frac{m_{P}^{2}}{2\pi er^{4}} 
\label{3.2}
\end{equation}
one finds that the energy density appears to be of gravitational origin, with $m_{P}/\sqrt{e}$ as the mass source. Moreover, the stress tensor (3.1) resembles that of an electrostatic field for a point charge at rest, with the charge $q = 2m_{P}/\sqrt{e}$. We would like to stress that all the components of $T^{a}_{~b}$ are finite everywhere, a property valid for all curvature invariants as well. The fact that the tangential pressures are negative for $r < r_{0}/2$ leads to a violation of the WEC (the condition $\rho > |p_{\theta}|$ is not obeyed for $r < r_{0}/4$). 

Let us compute the energy density $\rho$ taking an atom as source of the field, with $m \approx 10^{-24}$ g, at $r = 10^{-7}$ cm from its center. We have $2\hbar/ecmr \approx 10^{-6}$ and the exponential factor is practically unity. One obtains $\rho \approx \hbar c/2\pi er^{4} \approx 10^{11} erg/cm^{3}$.

\section{Komar energy}
 Being directly related to the radial acceleration, we prefer to use the Tolman-Komar expression for the quasilocal energy of our stress tensor (3.1)
   \begin{equation}
W = 2 \int(T_{ab} - \frac{1}{2} g_{ab}T^{c}_{~c})u^{a} u^{b} N\sqrt{\gamma} d^{3}x ,
\label{4.1}
\end{equation}  
which is measured by a static observer. $N$ in (4.1) is the lapse function and $\gamma$ is the determinant of the spatial 3-metric. The details of the calculations are similar with those from \cite{HC1, HC4} and, therefore, we get
 \begin{equation}
 W = m\left(1 - \frac{r_{0}}{r}\right) e^{-\frac{r_{0}}{r}},
\label{4.2}
\end{equation}
which may also be written as $W = r^{2} a^{r}$. It is worth noting that the expression (4.2) coincides with the ADM quasilocal energy function $E(r)$ from \cite{SC}. The ADM energy may be obtained from (4.2) when $r \rightarrow \infty$. We have $W \rightarrow 0$ when $r \rightarrow 0$ and $W \rightarrow m$ at infinity. If all fundamental constants are introduced in (4.2), we get
 \begin{equation}
 W = \left(mc^{2} - \frac{2\hbar c}{er}\right) e^{-\frac{2\hbar}{emcr}}.
\label{4.3}
\end{equation}
The case $\hbar = 0$ leads to the standard result $W = mc^{2}$. We see from (4.3) how the classical and quantum terms, $ mc^{2}$ and, respectively, $2\hbar c/er$, bring their contribution at the expression of $W(r)$. For $r >> r_{0}$, the classical term dominates. At $r = r_{0}$, $W$ vanishes but it becomes negative for $r < r_{0}$, with $W_{min} = -m/e^{2}$ at $r = r_{0}/2$. This seems to be a steady state which is rooted from the negative pressures contribution. 

As we have seen, for $m \geq m_{P}$, our particle may become a BH. We could compute the entropy $S_{H}$ of the BH horizon, following Padmanabhan's prescription \cite{TP} (see also \cite{HC1}). Let us restrict to the extremal situation when $m = m_{P}$ and $r_{H} = 2/e m_{P} = 2l_{P}/e$. We have
\begin{equation}
S_{H} = \left(\frac{|W|}{2T}\right)_{H} = \frac{4\pi}{e^{2}} = \frac{A_{H}}{4}
\label{4.4}
\end{equation}
which is of the order of unity. $T$ in (4.4) is the BH temperature and $A_{H}$ is the horizon area. Although $W$ and $T$ vanish separately at the horizon, their ratio is finite, leading to a finite horizon entropy.

\section{A version with time dependent mass}
As we know, a time dependent source with spherical symmetry will no longer lead to a Ricci-flat geometry, i.e., to a vacuum solution of the Einstein equations. Therefore, Birkhoff's theorem does not apply for this case. The line-element will be given by
  \begin{equation}
  ds^{2} = -\left(1- \frac{2m(t)}{r}\right) dt^{2} + \frac{dr^{2}}{1-\frac{2m(t)}{r}} dr^{2} + r^{2} d \Omega^{2}, 
 \label{5.1}
 \end{equation}
where $d \Omega^{2}$ stands for the metric on the unit two-sphere and $m(t)$ is the time dependent mass source. It is worth noticing that the source of the geometry (5.1) is an anisotropic fluid with zero energy density $\rho$ and zero radial pressure $p_{r}$. In contrast, the fluid has nonzero tangential pressures $p_{t}$ given by
  \begin{equation}
	p_{t} = T^{\theta}_{~\theta} = T^{\phi}_{~\phi} = \frac{2m(t) \ddot{m}(t) - 4\dot{m}^{2}(t) - r\ddot{m}(t)}{8\pi r^{2}\left(1 - \frac{2m(t)}{r}\right)^{3}},
 \label{5.2}
 \end{equation}
where $\dot{m}(t) = dm(t)/dt$, etc. In addition, there is an energy flux on the radial direction given by
  \begin{equation}
	 T^{r}_{~t} = \frac{\dot{m}(t)}{4\pi r^{2}}.
 \label{5.3}
 \end{equation}
Both $p_{t}$ and $T^{r}_{~t}$ turn out to become divergent when $r \rightarrow 2m(t)$ and $r \rightarrow 0$, respectively, for constant time. We shall see later how that inconvenient property will be removed, with an appropriate choice of the function $m(t)$. Moreover, we may get rid of the divergence of the scalar curvature
  \begin{equation}
	 R^{a}_{~a} = -2\frac{2m(t) \ddot{m}(t) - 4\dot{m}^{2}(t) - r\ddot{m}(t)}{r^{2}\left(1 - \frac{2m(t)}{r}\right)^{3}},~~~a = 0,1,2,3
 \label{5.4}
 \end{equation}
at the apparent horizon $r_{AH} = 2m(t)$, by means of the same recipe. We also observe that the acceleration 4-vector of a static observer in the spacetime (2.1) has only one nonzero component, $a^{r} = m(t)/r^{2}$.

Motivated by a recent paper \cite{HC2}, we propose the following geometry outside a variable mass with spherical symmetry
  \begin{equation}
  ds^{2} = -\left(1- \frac{2m}{r} e^{-\frac{k}{t}}\right) dt^{2} + \frac{dr^{2}}{1-\frac{2m}{r}e^{-\frac{k}{t}}} dr^{2} + r^{2} d \Omega^{2}, 
 \label{5.5}
 \end{equation}
with $lne = 1,~m(t) = me^{-\frac{k}{t}},~t>0$ and $m,k$ - positive constants. For a macroscopic $m$, we choose $k = 2m$, and the value $k = 1/m$ is selected for a microscopic particle . From now on we will deal with the macroscopic situation only, so that (5.5) yields
  \begin{equation}
  ds^{2} = -\left(1- \frac{2m}{r} e^{-\frac{2m}{t}}\right) dt^{2} + \frac{dr^{2}}{1-\frac{2m}{r}e^{-\frac{2m}{t}}} dr^{2} + r^{2} d \Omega^{2}. 
 \label{5.6}
 \end{equation}
To avoid a signature flip for the metric coefficients $g_{tt},~g_{rr}$, we impose the condition $f(r,t) \equiv 1- \frac{2m}{r} e^{-\frac{2m}{t}} >0$, namely $r > 2m e^{-\frac{2m}{t}}$, with $r_{AH} = 2m e^{-\frac{2m}{t}}$ - the location of the apparent horizon. 

For constant $r$, $f(r,t)$ is a monotonic decreasing function of $t$, tends to unity when $t \rightarrow 0$ and acquires the standard Schwarzschild value $(1 - 2m/r)$ at infinity (or when $t >> 2m$). When $f(r,t)$ is considered as a function of $r$, it equals unity for $r \rightarrow \infty$. However, the limit $r \rightarrow 0$ has to be taken with $t \rightarrow 0$, in order to satisfy the condition $r > 2m(t)$. Consequently, $0 < f(r,t) <1$. We notice also that the apparent horizon is an increasing function of $t$, from $r_{AH} \rightarrow 0$ when $t \rightarrow 0$ and $r_{AH} \rightarrow 2m$ at infinity, having an inflexion point at $t = m$. 

 The time variable $t$ in (5.6) is conjectured to represent, in our view, the duration of a measurement performed on a physical system located at some $r$, in the spacetime (5.6). For example, in the gravitational field of the Earth, with $M = 6 \cdot 10^{27} g$ and $2GM/c^{2} \approx 1 mm$, one obtains $2M/t = 2GM/c^{3}t = 3 \cdot 10^{-12}/t$. For a measurement done in $t = 10^{-12}$s, the exponential factor becomes $e^{-3}$, as if the Earth mass dropped $e^{3}$ times. That effect might be observed experimentally, if we consider, for instance, a particle (say, a proton) from the cosmic rays traveling to the Earth. Its radial acceleration w.r.t. an apparatus on the surface is given by $a^{r} = (M/R^{2})~exp(-2M/t)$, where $M$ and $R$ are, respectively, the Earth mass and radius. Taking the same value of the duration of measurement, we find that $ a^{r} = g/e^{3} \approx g/20$, where $g = 9.81 m/s^{2}$, so that the particle motion is slowing down. We took into consideration a cosmic proton for to get a measurable effect (the particle will travel a visible distance in $10^{-12}$ s because of its huge velocity).    

If the above model is valid, it represents an effective way to diminish the influence of gravity on a physical system. From a different point of view, Diosi \cite{LD2} presented a similar effect for a static source in the Newtonian gravitational field of the Earth, when it is shifted upward by a universal height $\delta = g\tau^{2}$, where $\tau \approx 1 ms$ is the delay time \cite{LD2} (see also \cite{RP, LD1, DB}). 

We might now give a plausible explanation to the fact that the zero point energy does not gravitate: the very fast quantum vacuum fluctuations reduce the strength of gravity so much that its influence is canceled. A similar phenomenon takes place for a non-traversable wormhole: the time variation of the wormhole throat radius is so fast that even light is not able to pass through.

Let us see now what are the properties of the anisotropic gravitational fluid.
As the line element (5.6) is not Ricci-flat, a source stress tensor is necessary on the r.h.s. of Einstein's equations $G_{ab} = 8\pi T_{ab}$, for to have (5.6) as an exact solution. The source is an anisotropic fluid with the only nonzero components  
  \begin{equation}
	 T^{r}_{~t} = \frac{m^{2}e^{-\frac{2m}{t}}}{2\pi r^{2}t^{2}},~~~T^{t}_{~r} = -\frac{m^{2}e^{-\frac{2m}{t}}}{2\pi r^{2}t^{2}\left(1- \frac{2m}{r} e^{-\frac{2m}{t}}\right)^{2}}
 \label{5.7}
 \end{equation}
and
  \begin{equation}
	p_{t} = \frac{m^{2}e^{-\frac{2m}{t}}}{2\pi rt^{3}\left(1- \frac{2m}{r} e^{-\frac{2m}{t}}\right)^{2}}\left[1 - \frac{m\left(1 + \frac{2m}{r}e^{-\frac{2m}{t}}\right)}{t\left(1 - \frac{2m}{r}e^{-\frac{2m}{t}}\right)}\right]
 \label{5.8}
 \end{equation}
The above fluid is ''exotic'', with zero energy density $\rho$, zero radial pressure $p_{r}$ but nonzero transversal pressures (as if the fluid were located in very thin spherical layers). The fact that $\rho$ is vanishing is understandable, otherwise the Equivalence Principle will be violated (there is no a local definition of the gravitational energy).

One observes that all components of $T^{a}_{~b}$ vanish when $t \rightarrow \infty$ (or when $t>>2m$) because the metric (5.6) becomes Ricci-flat. We must remind that the limit $r \rightarrow 0$ goes simultaneously with $t \rightarrow 0$ so that $T^{a}_{~b}$ tends to zero at this limit, too. That takes place because of the exponential factor $e^{-\frac{2m}{t}}$ which is present in all expressions, including the scalar curvature $R^{a}_{~a} = -8\pi T^{a}_{~a} = -16\pi p_{t}$. Moreover, in the latter case ($t \rightarrow 0$), the geometry (5.6) becomes Minkowskian and the effective mass $m(t)$ cancels. As far as the transversal pressure is concerned, we note that it vanishes at
 \begin{equation}
r_{0} = \frac{t+m}{t-m}\cdot 2m e^{-\frac{2m}{t}} > 2m e^{-\frac{2m}{t}},~~~t>m,
 \label{5.9}
 \end{equation}
which becomes $r_{0} \approx 2m$ for $t >> m$. One finds that $p_{t} < 0$ for $0<t<m$ and any $r$. In contrast, for $t>m$, we have $p_{t} \geq 0$, if $r \geq r_{0}$ and $p_{t} < 0$ if $r < r_{0}$.

 The acceleration 4-vector $a^{b}$ of a ''static'' observer in the geometry (5.6) has only one nonzero component
     \begin{equation}
	 a^{r} = \frac{m}{r^{2}} e^{-\frac{2m}{t}},
 \label{5.10}
 \end{equation}
that becomes the Newtonian expression for $t >> 2m$. The other limit ($t << 2m$) gives us $ a^{r} \approx 0$, as expected. In terms of $t$, at constant $r$, $a^{r}$ is a monotonically  increasing function. In addition, it vanishes when $r \rightarrow 0$, for the same reasons given before.

Our next task is to compute the Brown - York quasilocal energy corresponding to our spacetime (5.6).
Having now the components of the stress tensor and the basic physical quantities associated to it, our next task is to compute the total energy flow measured by an observer laying at r = const. \cite{HC5}
 \begin{equation}
 E = \int{T^{a}_{~b}u^{b}n_{a}\sqrt{-\gamma}}dt~d\theta~d\phi,
 \label{5.11}
 \end{equation}
where $n_{a}$ is a unit spacelike vector orthogonal to $u^{a}$, namely $n_{a} = (0, 1/\sqrt{1- \frac{2m(t)}{r}}, 0, 0)$ and $\gamma = -(1 - 2m(t)/r)r^{4} sin^{2}\theta$. With $T^{r}_{~t}$ from (5.7), Eq. (5.11) gives us
 \begin{equation}
E = \int{\frac{\dot{m}(t) dt}{\sqrt{1 - \frac{2m(t)}{r}}}},
 \label{5.12}
 \end{equation}
with $r$ fixed and $m(t)$ as the variable of integration. One obtains
 \begin{equation}
 E(m(t)) = -r \sqrt{1 - \frac{2m(t)}{r}} +h(r),
 \label{5.13}
 \end{equation}
where $h(r)$ is a constant of integration. It may be found by imposing that $E = 0$ when $m(t)$ vanishes, so that $h(r) = r$. Hence
 \begin{equation}
 E(r,t) = r \left(1 -  \sqrt{1 - \frac{2m}{r}e^{-\frac{2m}{t}}}\right).
 \label{5.14}
 \end{equation}
 We might, of course, calculate $E(r,t)$ directly, following the recipe from \cite{LSY}
 \begin{equation}
 E(r,t) = \frac{1}{8\pi} \int_{B}{(K - K_{0})\sqrt{\sigma}d^{2}x},
  \label{5.15}
	\end{equation}
where $\sigma = det(\sigma_{ab})$ and $\sigma_{ab} = g_{ab} + u_{a}u_{b} - n_{a}n_{b}$ is the induced metric on the two boundary $B$. $K$ in (5.15) is the trace of the extrinsic curvature of $B$ and $K_{0}$ corresponds to the vacuum (i.e., when $m = 0$). But the connection coefficients $\Gamma_{r\theta}^{\theta}$ and $\Gamma_{r\phi}^{\phi}$ for the metric (5.6), that are needed to compute $K$ from (5.15), are the same as in the static case so that the expression (5.14) emerges.

One notices that $E$ tends to zero when $t \rightarrow 0$ and acquires the Brown-York (static) form $E_{BY} = r(1 -  \sqrt{1 - \frac{2m}{r}})$ when $t >> 2m$. For constant $r$, $E$ is a monotonic increasing function of $t$, with $0 < E < E_{BY}$. We see there is no energy flux when $t \rightarrow 0$, whatever the value of $r$. That is a consequence of the flatness of (5.6) in that limit.

\section{Conclusions}
The problem whether the evaporation of a BH stops and a remnant arises sparked much interest in the last decade. A remnant is formed when gravity becomes repulsive at very short distances, as it was suggested in this work. As several previous authors, we proposed a modified version of the KS geometry, to fit better in microsystems. The source of the curvature is an anisotropic fluid with $p_{r} = -\rho$ and which depends on $\hbar$ and $c$ only, but not on the Newton constant $G$. We conclude that a particle of mass $m$ may become a BH when $m \geq m_{P}$ only, otherwise the red-shift factor has no zeros. The Planck constant $\hbar$ plays a very important role in our investigations because the role of the gravitational radius is replaced with the Compton wavelength associated to the source $m$.

We also investigated the role of the measurement process in gravitational physics. Our research is based on the papers by Rovelli \cite{CR}, Hohn \cite{PH} and Okon and Sudarsky \cite{OS}, who suggest that there are no observer independent values of physical quantities. Reality, rather than being a given entity,  is just an interpretation of our interactions with the world around us.

In the time dependent version we have proposed, the time variable plays the role of the duration of measurement upon some physical system. Very short time intervals lead to much weaker values of the gravitational field where our system is located. That may direct us to an explanation of the well-known fact that the vacuum energy does not gravitate: very fast quantum fluctuations get rid of the influence of gravity.

\end{document}